\pdfoutput=1

\documentclass[pra,onecolumn,showpacs,longbibliography]{revtex4-1}

\usepackage{amssymb}
\usepackage{amsmath}
\usepackage{graphicx}
\usepackage{color}
\usepackage{siunitx}
\usepackage{grffile}

\newcommand{\imu}{\mathrm{i}}
\newcommand{\dint}[2][]{\!\mathrm{d}^{#1}#2\,}                 
\newcommand{\dfint}[3][]{\! \frac{\mathrm{d}^{#1}#2}{#3}\,}
\renewcommand{\Re}{\mathrm{Re}\,}
\renewcommand{\Im}{\mathrm{Im}\,}
\renewcommand{\vec}[1]{\mathbf{{#1}}}                         
\newcommand{\pvec}[1]{\mathbf{{#1}}_\parallel }
                         
\newcommand{\pvecUnit}[1]{\mathbf{\hat{#1}}_\parallel }
\newcommand{\ooc}[1]{\frac{\omega^{#1}}{c^{#1}}}

\graphicspath{{Figures/},.}         

\begin{document}

\title{Features in the diffraction of a scalar plane wave from doubly-periodic Dirichlet and Neumann surfaces}

\author{Alexei A. Maradudin}
\affiliation{Department of Physics and Astronomy, University of California, Irvine CA 92697, U.S.A.}
\email{aamaradu@uci.edu}

\author{Veronica P\'erez-Ch\'avez}
\affiliation{Centro de Ense\~nanza T\'ecnica y Superior, Universidad Ensenada
Camino a Microondas Trinidad s/n Km.~1, Moderna Oeste, 22860 Ensenada, B.C., M\'exico.}

\author{Arkadiusz J\c{e}drzejewski}
\affiliation{Department of Theoretical Physics, Wroclaw University of Technology, Wroclaw, Poland}

\author{Ingve Simonsen}
\affiliation{Surface du Verre et Interfaces, UMR 125 CNRS/Saint-Gobain, F-93303 Aubervilliers, France}
\affiliation{Department of Physics, NTNU --- Norwegian University of Science and Technology, NO-7491 Trondheim, Norway}
\affiliation{Department of Petroleum Engineering, University of Stavanger, NO-4036 Stavanger, Norway}
\email{Ingve.Simonsen@ntnu.no}

\begin{abstract}
The diffraction of a scalar plane wave from a doubly-periodic  surface on which either the Dirichlet or Neumann boundary condition is imposed is studied by means of a rigorous numerical solution of the Rayleigh equation for the amplitudes of the diffracted Bragg beams. From the results of these calculations the diffraction efficiencies of several of the lowest order diffracted beams are calculated as functions of the polar and azimuthal angles of incidence. The angular dependencies of the diffraction efficiencies display features that can be identified as Rayleigh anomalies for both types of surfaces. In the case of a Neumann surface additional  features are present that can be attributed to the existence of surface waves on such surfaces. Some of the results obtained through the use of the Rayleigh equation are validated by comparing them with results of a rigorous Green's function numerical calculation.  
\end{abstract}

\maketitle


\section{Introduction}

In several resent papers the present authors have studied theoretically and computationally the diffraction of p-polarized light from a perfectly conducting grating~\cite{23} and from a high-index dielectric grating~\cite{N2}, and the diffraction of a shear horizontally polarized acoustic wave from a gating ruled on the surface of an isotropic elastic medium~\cite{N3}. Each of these media does not support a surface wave when the surface bounding it is planar, but supports one when it is periodically corrugated. The dependence of the diffraction efficiencies of some of the lowest-order Bragg beams on the angle of incidence was found to posses two types of anomalies. The first type of anomaly occurred at angles of incidence at which a diffractive order begins to propagate or ceases to propagate. They were first observed by Wood in 1902~\cite{15} in the diffraction of light from a metallic grating. Their origin was explained by Lord~Rayleigh~\cite{16}, and they are now called \textit{Rayleigh anomalies}. The second type of anomaly occurred at angles of incidence at which the incident field excites a surface wave supported by the grating, when the difference of the components of the wave vectors of the incident field parallel to the mean scattering plane and of the surface wave is made up by the addition of a grating vector. These anomalies were also first observed by Wood in 1902~\cite{15}, and were shown to be due to the grating-induced excitation of surface plasmon polaritons by the incident light by Fano~\cite{22}. We refer to them as \textit{Wood anomalies}. Since surface waves don't exist on planar surfaces of the media studied in Refs.~\cite{23,N2,N3}, our results obtained in these papers emphasized the necessity of surface waves for the existence of Wood anomalies, and that the surface waves need not be surface plasmon polaritons, but can be of a quite different nature.

In this paper we extend the work presented in Refs.~\cite{23,N2,N3} to the case of diffraction of a scalar plane wave from a doubly-periodic grating (often called a bigrating or a cross-grating), fabricated on a surface of the medium that when planar does not support a surface wave, but can support one when it is doubly periodic. Specifically, we consider the diffraction of a scalar plane wave from a doubly-periodic surface on which either the Dirichlet or the Neumann boundary condition is satisfied. For brevity, we will refer to these two types of surfaces as Dirichlet and Neumann surfaces, respectively. It is known that a doubly-periodic Neumann surface supports a surface wave while a doubly-periodic Dirichlet surface does not~\cite{14}. We calculate the dependence of several of the lowest-order diffraction efficiencies on the polar angle of incidence for a fixed azimuthal angle of incidence, and look for Rayleigh anomalies in these dependencies for both types of surfaces, and for Wood anomalies in the diffraction from a Neumann surface.

Calculations of the dependence of the efficiencies of diffraction of light from, or its transmission through, several types of doubly-periodic structures on the wavelength of the incident light, or on its polar angle of incidence for a fixed azimuthal angle of incidence, have been carried out by a variety of approaches. Some of them have been devoted to the wavelength dependence of the reflectivity and the dip it displays that arises from the excitation of a surface wave supported by the doubly-periodic structure~\cite{N8,N9,N10,N11}, others to the phenomenon of total absorption~\cite{N12}, and still others to the wavelength dependence of the enhanced transmission of light through a doubly-periodic array of nanoscale holes piercing a thin metal film~\cite{N13,N14,N15,N16}. Similar calculations of higher-order diffraction efficiencies, such as the ones carried out in the present work, do not appear to have been carried out in these earlier studies.

The  calculations in the body of this paper will be carried out on the basis of the Rayleigh hypothesis~\cite{1}, perhaps the simplest approach to solving this scattering problem. The validity of this approach has been  questioned on occasion~\cite{2,3,4}, primarily on the grounds that if the indentations of the surface are sufficiently deep some of the scattered waves can be propagating downward toward the surface within them, before becoming upward outgoing propagating waves due to multiple scattering. Such waves are not taken into account by the Rayleigh hypothesis, which assumes only upward outgoing propagating waves everywhere above the surface. Nevertheless, in subsequent work limits of validity of this hypothesis have been determined for the scattering of a scalar plane wave from a singly-periodic surface~\cite{5,6,7,8,9,10} and from a doubly-periodic surface~\cite{11}, defined by profile functions that are analytic functions of the coordinates in the mean scattering plane. It has recently been argued that the Rayleigh hypothesis is always valid~\cite{12}.

With this background, in this paper we derive the Rayleigh equations for the diffraction of a scalar plane wave from doubly-periodic Dirichlet and Neumann surfaces, and solve them numerically. For greater generality, and for pedagogical reasons, we begin the derivation by first obtaining the Rayleigh equation for the scattering of a scalar plane wave from an arbitrary rough two-dimensional Dirichlet and Neumann surface, and then specialize this equation to the case of a doubly-periodic surface. From the solutions of these equations we will calculate the angular dependencies of several of the diffraction orders. The validity of the Rayleigh hypothesis in the context of the problem studied will be demonstrated by a comparison of some of the results obtained by its use with those obtained by a rigorous numerical method~\cite{17,17a}.

\section{Scattering Theory}

The system that we consider consists of a liquid in the region $x_3>\zeta(\pvec{x})$ and an impenetrable medium in the region $x_3<\zeta(\pvec{x})$ where $\pvec{x}=(x_1,x_2,0)$ is a position vector in the plane $x_3=0$. The surface profile function $\zeta(\pvec{x})$ is assumed to be a single valued function of $\pvec{x}$, and to be differentiable with respect to $x_1$ and $x_2$.

The field $\psi(\vec{x};t)$ in the region $x_3>\zeta(\pvec{x})$ consists of an incoming incident scalar plane wave of frequency $\omega$ and a superposition of outgoing scattered plane waves of the same frequency $\psi(\vec{x};t)=\left[\psi(\vec{x}|\omega)_{\mathrm{inc}} + \psi(\vec{x}|\omega)_{\mathrm{sc}}\right]
\exp(-\imu\omega t) \equiv \psi(\vec{x}|\omega)\exp(-\imu\omega t)$. The amplitude function $\psi(\vec{x}|\omega)$ is the solution of the Helmholtz equation
\begin{align}
  \left[\nabla^2 +\ooc{2} \right] \psi(\vec{x}|\omega) &= 0,
  \label{eq:1}
\end{align}
where $c$ is the speed of the field in the liquid. The field satisfies either (a)~the Dirichlet boundary condition
\begin{align}
  \label{eq:2}
  \psi(\vec{x}|\omega)\big|_{x_3=\zeta(\pvec{x})}  &= 0    
\end{align}
or (b)~the Neumann boundary condition
\begin{align}
  \label{eq:3}
  \frac{\partial \psi(\vec{x}|\omega)}{\partial n}\Big|_{x_3=\zeta(\pvec{x})} &= 0 
\end{align}
on the rough surface $x_3=\zeta(\pvec{x})$. In Eq.~\eqref{eq:3} $\partial/\partial n$ is the derivative along the normal to the surface at each point of it  directed into the region $x_3>\zeta(\pvec{x})$,
\begin{align}
  \label{eq:4}
   \frac{\partial\;\;}{\partial n}
  &=
    \frac{1}{ \left[1+\left\{\nabla\zeta(\pvec{x})\right\}^2 \right]^{1/2}}
    \left[
    -\zeta_1(\pvec{x}) \frac{\partial\quad\! }{\partial x_1}
    -\zeta_2(\pvec{x}) \frac{\partial\quad\!}{\partial x_2}
    +                  \frac{\partial\quad\!}{\partial x_3} 
    \right],
\end{align}
where $\zeta_\alpha(\pvec{x}) =\partial \zeta(\pvec{x})/\partial x_\alpha$ ($\alpha=1,2$). It is clear that the prefactor $[1+\{\nabla\zeta(\pvec{x})\}^2 ]^{-1/2}$ on the right-hand side of Eq.~\eqref{eq:4} can be neglected in what follows.

In the scattering of a scalar plane wave from either type of surface, the incident field $\psi(\vec{x}|\omega)_{\mathrm{inc}}$  can be written as
\begin{align}
  \psi(\vec{x}|\omega)_\mathrm{inc}
  &= \exp\left[ \imu\pvec{k}\cdot\pvec{x} - \imu \alpha_0(k_\parallel,\omega) x_3  \right],
    \label{eq:5}
\end{align}
where $\pvec{k}=(k_1,k_2,0)$ and $\alpha_0(k_\parallel,\omega)=\left[(\omega^2/c^2)-k_\parallel^2 \right]^{1/2}$, with $k_\parallel < \omega/c$.

Similarly,  the field scattered from either type of surface, $\psi(\vec{x}|\omega)_{\mathrm{sc}}$, can be written
 \begin{align}
  \psi(\vec{x} | \omega)_\mathrm{sc}
  &=
    \int \dfint[2]{q_\parallel}{{(2\pi)^2}}
    R(\pvec{q}|\pvec{k})
    \exp
    \left[
    \imu\pvec{q}\cdot\pvec{x} + \imu \alpha_0(q_\parallel,\omega) x_3
    \right],
     \label{eq:6}
\end{align}
where $\alpha_0(q_\parallel, \omega) =  \left[(\omega^2/c^2)-q_\parallel^2\right]^{1/2}$ with $\Re \alpha_0(q_\parallel, \omega)>0$ and $\Im \alpha_0(q_\parallel, \omega) > 0$. Of course the scattering amplitude $R(\pvec{q}|\pvec{k})$ will be different for the scattering from a Dirichlet surface  then it is for scattering from a Neumann surface due to the different boundary conditions satisfied on the two types of surfaces.

We now substitute the sum of Eqs.~\eqref{eq:5} and \eqref{eq:6} into the boundary conditions~\eqref{eq:2} and \eqref{eq:3}. This step represents our assumption of the validity of the Rayleigh hypothesis~\cite{1}. The resulting equations for the scattering amplitude can be written as
\begin{align}
  &\begin{pmatrix}
    1
    \\
    -\left[
      \pvec{k}\cdot\vec{\nabla}\zeta(\pvec{x}) + \alpha_0(k_\parallel,\omega)
    \right]
  \end{pmatrix}
  \exp
  \left[
  \imu \pvec{k}\cdot\pvec{x} - \imu \alpha_0(k_\parallel,\omega) x_3
  \right]
  \nonumber
  \\ & \qquad \qquad  
  +
  \int \dfint[2]{q_\parallel}{{(2\pi)^2}} 
  R(\pvec{q}|\pvec{k})
  \begin{pmatrix}
    1
    \\
    -\left[
      \pvec{q}\cdot\vec{\nabla}\zeta(\pvec{x}) - \alpha_0(q_\parallel,\omega)
    \right]
  \end{pmatrix}
  \exp
  \left[
  \imu \pvec{q}\cdot\pvec{x} + \imu \alpha_0(q_\parallel,\omega) x_3
  \right]
=
  0.
  \label{eq:7}
\end{align}

We now introduce the function $I(\gamma|\pvec{Q})$ by
\begin{subequations}
  \label{eq:8}
\begin{align}
  \exp\left[ -\imu \gamma\zeta(\pvec{x}) \right]
  &=
    \int  \dfint[2]{Q_\parallel}{(2\pi)^2} 
    I(\gamma | \pvec{Q}) \exp\left[ \imu \pvec{Q} \cdot \pvec{x}\right],
  \label{eq:8a}
\end{align}
so that
\begin{align}
  I(\gamma | \pvec{Q}) 
  &=
    \int \dint[2]{x_\parallel} 
    \exp\left[ -\imu \gamma\zeta(\pvec{x}) \right]
    \exp\left[ -\imu \pvec{Q} \cdot \pvec{x}\right].
  \label{eq:8b}
\end{align}
If we differentiate both sides of Eq.~\eqref{eq:8a} with respect to $x_\alpha$ ($\alpha=1,2$), we obtain the useful result
\begin{align}
  \zeta_\alpha(\pvec{x}) \exp\left[ -\imu \gamma\zeta(\pvec{x}) \right]
  &=
    - \int \dfint[2]{Q_\parallel}{(2\pi)^2}
    \frac{Q_\alpha}{\gamma}
    I(\gamma | \pvec{Q})
    \exp\left[ \imu \pvec{Q} \cdot \pvec{x}\right].
\end{align}
\end{subequations}

When we substitute Eqs.~\eqref{eq:8} into Eqs.~\eqref{eq:7} and equate to zero the coefficient of $\exp[\imu\pvec{p}\cdot\pvec{x}]$ in the resulting equations, the equations satisfied by the scattering amplitudes $R(\pvec{q}|\pvec{k})$ become
\begin{align}
  \int \dfint[2]{q_\parallel}{(2\pi)^2} 
  I\left( -\alpha_0(q_\parallel,\omega) \big| \pvec{p}-\pvec{q} \right)
  M(\pvec{p}|\pvec{q})
  R(\pvec{q}|\pvec{k})
  &=
    -
    I\left( \alpha_0(k_\parallel,\omega) \big| \pvec{p}-\pvec{k} \right)
    N(\pvec{p}|\pvec{k}),   
  \label{eq:9}
\end{align}
where
\begin{align}
  M(\pvec{p}|\pvec{q}) &= 1,    &   N(\pvec{p}|\pvec{k}) &= 1
  \label{eq:10}                                            
\end{align}
for a Dirichlet surface, and
\begin{align}
  M(\pvec{p}|\pvec{q})
  &=
    \frac{ (\omega/c)^2
    - \pvec{p} \cdot \pvec{q}
    }{
    \alpha_0(q_\parallel,\omega)
    },
  &
  N(\pvec{p}|\pvec{k})
  &=
    -
    \frac{
    (\omega/c)^2
    - \pvec{p} \cdot \pvec{k}
    }{
    \alpha_0(k_\parallel,\omega).
    }
  \label{eq:11}
\end{align}
for a Neumann surface. Equations~\eqref{eq:9}--\eqref{eq:11} constitute the Rayleigh equations for the scattering amplitude in the scattering of a scalar plane wave from a two-dimensional rough Dirichlet or Neumann surface.

\section{The Differential Reflection Coefficient}

The scattering amplitude $R(\pvec{q}|\pvec{k})$ is of great importance in calculations of scattering from rough surfaces because an experimentally accessible quantity, the differential reflection coefficient, is expressed in terms of it.
The differential reflection coefficient $\partial R/\partial\Omega_s$ is defined such that $(\partial R/\partial\Omega_s)\,\dint\Omega_s$ is the fraction of the total time averaged incident flux that is scattered into an element of solid angle $\dint\Omega_s$ around the direction defined by the polar and azimuthal angles of scattering $(\theta_s,\phi_s)$.

The magnitude of the total time-averaged flux incident on the surface is given by 
\begin{align}
  P_{\mathrm{inc}} 
  &=  
    -
    A\,
    \Im\!\!
    \int\limits_{-\frac{L_1}{2}}^{\frac{L_1}{2}} \!\! \dint{x_1}
    \int\limits_{-\frac{L_2}{2}}^{\frac{L_2}{2}} \!\! \dint{x_2}
    \left[ 
    \psi^*(\vec{x}|\omega)_{\mathrm{inc}}
    \frac{\partial \psi(\vec{x}|\omega)_{\mathrm{inc}}}{\partial x_3}
    \right]_{x_3>\max\zeta(\pvec{x})},
  \label{eq:12}
\end{align}
where $L_1$ and $L_2$ are the lengths of the scattering surface along the $x_1$ and $x_2$ axes, while $A$ is a coefficient that drops out of the expression for the differential reflection coefficient. (For the scattering of a particle of mass $m$, $A=\hbar/m$ where $\hbar$ denotes Planck's constant.). The minus sign that appears in on the right-hand side of Eq.~\eqref{eq:12} compensates for the fact that the incident flux is negative. For the form of the incident field biven by Eq.~\eqref{eq:5} we find easily that
\begin{align}
  P_{\mathrm{inc}}
  &=
    AL_1L_2 \alpha_0(k_\parallel,\omega), 
  \label{eq:13}
\end{align}
where we have used the fact that $\alpha_0(k_\parallel,\omega)$ is real.

Similarly, the magnitude of the total time-averaged scattered flux is given by
\begin{align}
  P_{\textrm{sc}} 
  &=  
    A \;
    \Im\!\!
    \int\limits_{-\frac{L_1}{2}}^{\frac{L_1}{2}} \!\! \dint{x_1}
    \int\limits_{-\frac{L_2}{2}}^{\frac{L_2}{2}} \!\! \dint{x_2}
    \left[ 
    \psi^*(\vec{x}|\omega)_{\textrm{sc}}
    \frac{\partial \psi(\vec{x}|\omega)_{\textrm{sc}}}{\partial x_3}
    \right]_{x_3>\max\zeta(\pvec{x})}.
  \label{eq:14}
\end{align}
With the use of the expression for $\psi(\vec{x}|\omega)_{\mathrm{sc}}$ given by Eq.~\eqref{eq:6}, this expression becomes
\begin{align}
  P_{\textrm{sc}} 
  &=  
    A \;
    \Im\!\!
    \int\limits_{-\frac{L_1}{2}}^{\frac{L_1}{2}} \!\! \dint{x_1}
    \int\limits_{-\frac{L_2}{2}}^{\frac{L_2}{2}} \!\! \dint{x_2}
    \int \dfint[2]{q_\parallel}{(2\pi)^2}
    \int \dfint[2]{q'_\parallel}{(2\pi)^2}
    \imu \alpha_0(q'_\parallel,\omega)
    R^*(\pvec{q}|\pvec{k}) R(\pvec{q}'|\pvec{k})
    \exp\left[ -\imu\left(\pvec{q}-\pvec{q}'\right)\cdot \pvec{x}\right]
    \notag \\
  & \qquad \qquad \qquad \qquad
    \qquad \qquad \qquad \qquad
    \qquad 
    \times
    \exp
    \left\{
    -\imu \left[
    \alpha_0^*(q_\parallel,\omega) - \alpha_0(q'_\parallel,\omega) 
    \right] x_3
    \right\}
    \notag \\
  &=
    A\; \Im \imu \!\!
    \int \dfint[2]{q_\parallel}{(2\pi)^2}
    \alpha_0(q_\parallel,\omega)
    \left| R(\pvec{q}|\pvec{k}) \right|^2
    \exp \left[ -2\Im\alpha_0(q_\parallel,\omega) x_3 \right]
    \notag \\
  &=
    A
    \int\limits_{q_\parallel<\omega/c}
    \dfint[2]{q_\parallel}{(2\pi)^2}
    \alpha_0(q_\parallel,\omega)
    \left| R(\pvec{q}|\pvec{k}) \right|^2.    
  \label{eq:15}
\end{align}
In obtaining this result we have used the fact that $\alpha_0(q_\parallel,\omega)$ is real for $0<q_\parallel<\omega/c$, while it is imaginary for $q_\parallel>\omega/c$, to obtain the domain of integration indicated.

We now introduce the polar and azimuthal angles of incidence $(\theta_0,\phi_0)$ and of scattering $(\theta_s,\phi_s)$, respectively, through the relations
\begin{subequations}
     \label{eq:16}
\begin{align}
  \pvec{k}
  &=
    \frac{\omega}{c}\sin\theta_0 ( \cos\phi_0, \sin\phi_0, 0)
   \label{eq:16a}
\end{align}
and
\begin{align}
  \pvec{q}
  &=
    \frac{\omega}{c}\sin\theta_s ( \cos\phi_s, \sin\phi_s, 0).
  \label{eq:16b}
\end{align}
\end{subequations}
It follows that $\alpha_0(k_\parallel,\omega) = (\omega/c)\cos\theta_0$, $\alpha_0(q_\parallel,\omega) = (\omega/c)\cos\theta_s$, and $\dint[2]{q_\parallel}=(\omega/c)^2\cos\theta_s\,\mathrm{d}\Omega_s$ where $\mathrm{d}\Omega_s$, the element of solid angle, is $\mathrm{d}\Omega_s=\sin\theta_s\,\mathrm{d}\theta_s\,\mathrm{d}\phi_s$.
With the use of these results the incident flux can be written as
\begin{align}
  \label{eq:17}
  P_{\mathrm{inc}} (\theta_0)
  &=
    AL_1L_2 \ooc{} \cos\theta_0, 
\end{align}
while the scattered flux becomes
\begin{align}
  \label{eq:18}
  P_{\textrm{sc}} 
  &=
    \int \dint{\Omega_s}
    p_{\mathrm{sc}} (\theta_s, \phi_s)
\end{align}
where
\begin{align}
  \label{eq:19}
  p_{\mathrm{sc}} (\theta_s, \phi_s)
  &=
    A \left( \frac{\omega}{2\pi c}\right)^2 \ooc{} \cos^2\theta_s
    \left| R(\pvec{q}|\pvec{k}) \right|^2.        
\end{align}
By definition the differential reflection coefficient is
\begin{align}
  \frac{\partial R }{ \partial \Omega_s}
  &=
    \frac{ p_{\mathrm{sc}} (\theta_s, \phi_s) }{P_{\mathrm{inc}} (\theta_0)}
    \notag \\
  &=
    \frac{1}{L_1 L_2}
    \left( \frac{\omega}{2\pi c}\right)^2
    \frac{ \cos^2\theta_s }{ \cos\theta_0 }
    \left| R(\pvec{q}|\pvec{k}) \right|^2,        
\end{align}
where $\pvec{k}$ and $\pvec{q}$ are defined by Eqs.~\eqref{eq:16a} and \eqref{eq:16b}, respectively.

%

\section{Doubly-Periodic Surface}

The results obtained in the preceding sections of this paper apply to an arbitrary two-dimensional rough surface defined by the single-valued surface profile function $\zeta(\pvec{x})$  that is differentiable with respect to $x_1$ and $x_2$. In this section we specialize these results to the case where the surface profile function is a doubly-periodic function of $\pvec{x}$, namely a bigrating. 

Thus the surface  profile function $\zeta(\pvec{x})$ is assumed to possess the property $\zeta \left( \pvec{x} +\pvec{x}( \ell ) \right)=\zeta( \pvec{x})$, where the vector $\pvec{x}( \ell )$ is a translation vector of a two-dimensional Bravais lattice~\cite{[{See, for example, }][{, p.~8.}]18}.
It is defined by 
\begin{align}
  \pvec{x}( \ell )
  &=
    \ell_{1}\vec{a}_{1}+ \ell_{2}\vec{a}_{2},
    \label{eq:21}
\end{align}
where $\vec{a}_1$ and $\vec{a}_2$ are the two noncolinear primitive translation vectors of the Bravais lattice, while $\ell_1$ and $\ell_2$ are any positive or negative integers or zero, which we denote collectively by $\ell$.
The area of a primitive unit cell of this lattice is $a_c=| \vec{a}_1 \times \vec{a}_2|$.

We also introduce the lattice reciprocal to the one defined by Eq.~\eqref{eq:21}. Its lattice sites are defined by the vectors
\begin{align}
  \label{eq:22}
  \pvec{G}(h)
  &=
    h_1 \vec{b}_1 + h_2 \vec{b}_2,
\end{align}
where the primitive translation vectors $\vec{b}_1$ and $\vec{b}_2$ are related to the primitive translation vectors of the direct lattice, $\vec{a}_1$ and $\vec{a}_2$, by $\vec{a}_{i}\cdot \vec{b}_{j}= 2\pi \delta_{ij}$, while $h_1$ and $h_2$ are any positive or negative integers or zero, which we denote collectively by $h$. 
 
We now proceed to transform the Rayleigh equation~\eqref{eq:9} for the scattering amplitude $R(\pvec{q}|\pvec{k})$ into the Rayleigh equation for the corresponding amplitude that arises in the diffraction of a scalar plane wave from an impenetrable bigrating.

Due to the periodicity of the surface profile function $\zeta(\pvec{x})$, the field in the region $x_3>\zeta(\pvec{x})$ must satisfy the Floquet-Bloch condition~\cite{19,20} 
\begin{align}
  \label{eq:23}
  \psi\!
  \left(
  \pvec{x}+\pvec{x}(\ell),
  x_{3}| \omega
  \right)
  &=
    \exp
    \left[
    \imu \pvec{k} \cdot \pvec{x}( \ell )
    \right]
    \psi\!
    \left( \pvec{x},x_{3}| \omega \right).
\end{align}
This condition is satisfied if we rewrite the scattering amplitude $R(\pvec{q}|\pvec{k})$ in the form
\begin{align}
  \label{eq:24}
  R( \pvec{q} | \pvec{k})
  &=
    \sum_{\mathbf{G}_{\parallel}}
    \left( 2\pi \right)^{2}
    \delta
    \left(
    \pvec{q} - \pvec{k} -\pvec{G}
    \right)
    r
    ( 
    \pvec{k} + \pvec{G} | \pvec{k}
    ).
\end{align}
In writing this equation we have replaced summation over $h$ by summation over $\pvec{G}$.

A second consequence of the periodicity of the surface profile function $\zeta(\pvec{x})$ is that the function $I(\gamma|\pvec{Q})$ defined by Eq.~\eqref{eq:8b} can now be written
\begin{align}
  \label{eq:25}
  I\left( \gamma | \mathbf{Q}_{\parallel }\right)
  &=
    \sum_{\ell}
    \int\limits_{a_c(\ell)}
    \dint[2]{x_\parallel} 
    \exp
     \left[
    -\imu \gamma \zeta( \pvec{x})
    \right]
    \exp
    \left[
     -\imu \pvec{Q} \cdot \pvec{x}
     \right],
\end{align}
where $a_c(\ell)$ is the area of the unit cell containing the translation vector $\pvec{x}(\ell)$. The change of variable $\pvec{x}=\pvec{x}(\ell)+\pvec{u}$, and the relation $\zeta(\pvec{x}+\pvec{x}(\ell))=\zeta(\pvec{x})$, yield the result 
\begin{align}
  \label{eq:26}
  I\left( \gamma | \mathbf{Q}_{\parallel }\right)
  &=
    \sum_{\ell}
    \exp
    \left[
      -\imu \pvec{Q} \cdot \pvec{x}(\ell)
    \right]
    \int\limits_{a_c}
    \dint[2]{u_\parallel} 
    \exp
     \left[
    -\imu \gamma \zeta( \pvec{u})
    \right]
    \exp
    \left[
     -\imu \pvec{Q} \cdot \pvec{u}
     \right].
\end{align}
The use of the relation~\cite{21}
\begin{align}
  \label{eq:27}
  \sum_{\ell}
  \exp
  \left[
  - \imu \pvec{Q} \cdot \pvec{x}(\ell)
  \right]
  &=
    \sum_{\pvec{G}}
    \frac{ (2\pi)^2 }{ a_c }
    \delta
    \left(
    \pvec{Q}
    -
    \pvec{G}
    \right)
\end{align}
in Eq.~\eqref{eq:26} yields the result
\begin{align}
  \label{eq:28}
  I\left( \gamma | \mathbf{Q}_{\parallel }\right)
  &=
    \sum_{\mathbf{G}_{\parallel }}
    \left( 2\pi \right)^{2}
    \delta\! \left( \pvec{Q} - \pvec{G} \right)
    \widehat{I}\left( \gamma | \pvec{G} \right),
\end{align}
where
\begin{align}
  \label{eq:29}
  \widehat{I}\left( \gamma \mid \pvec{G} \right)
  &=
    \frac{1}{a_{c}}
    \int\limits_{a_{c}} \dint[2]{x_\parallel} 
    \exp \left[ -\imu \gamma \zeta( \mathbf{x}_{\parallel }) \right]
    \exp \left[ -\imu \pvec{G} \cdot \pvec{x} \right].
\end{align}

When the results given by Eqs.~\eqref{eq:24} and \eqref{eq:28} are substituted into Eq.~\eqref{eq:9}, and the integration over $\pvec{q}$ is carried out, we obtain the equation
\begin{align}
  \label{eq:30}
  \sum_{\mathbf{K}_{\parallel }}
  \left( 2\pi \right)^{2}
  &
  \delta
  \left(
    \pvec{p} -\pvec{K}
    \right)
  \;    
  \sum_{\mathbf{K}_{\parallel }^{\prime }}\;
  \widehat{I}
    \left(
    -\alpha_{0}( K_\parallel',\omega )
    \big|
    \pvec{K} - \pvec{K}'
    \right)
    M
    \left(
    \pvec{K}
    |
    \pvec{K}'
    \right)
      r( \pvec{K}' | \pvec{k} )   
    \notag \\
  &=
    -\sum_{\mathbf{K}_{\parallel }}
    \left( 2\pi \right)^{2}
    \delta
    \left(
       \pvec{p} - \pvec{K}
    \right)
    \widehat{I}
    \left(
    \alpha_{0}( k_\parallel, \omega)
    \big| \pvec{K} - \pvec{k} 
    \right)
    N
    \left(
    \pvec{K}
     |
     \pvec{k}
     \right).
\end{align}
In writing this equation, to simplify the notation we have defined the two wave vectors
\begin{align}
  \label{eq:31}
    \pvec{K} &= \pvec{k} + \pvec{G}
  &
    \pvec{K}' &= \pvec{k} + \pvec{G}',  
\end{align}
and have replaced summation over $\pvec{G}$ and $\pvec{G}'$ by summation over $\pvec{K}$ and $\pvec{K}'$, respectively. On equating the coefficients of $\delta(\pvec{p}-\pvec{K})$ on both sides of Eq.~\eqref{eq:30} we obtain the Rayleigh equation satisfied by $r(\pvec{K}|\pvec{k})$
\begin{align}
  \label{eq:32}
  \sum_{\mathbf{K}_{\parallel }^{\prime }}\;
    \widehat{I}
    \left(
    -\alpha_{0}( K_\parallel', \omega )
    \big|
    \pvec{K} - \pvec{K}'
    \right)
    M( \pvec{K} | \pvec{K}' )
    \,
    r( \pvec{K}'| \pvec{k} )
    &=
    -
    \widehat{I}
    \left(
    \alpha_{0}( k_\parallel, \omega)
    \big| \pvec{K} - \pvec{k} 
    \right)
    N ( \pvec{K} | \pvec{k} ).
\end{align}
Equation~\eqref{eq:32} holds for all possible values of $\pvec{K}$ (or $\pvec{G}$), and hence it  represents a linear system of equations of infinite dimension. To be able to solve the system numerically, we need a system of finite dimension. This can be achieved by restricting the vectors $\pvec{G}(h)$ and $\pvec{G}'(h)$ to a domain for which their lengths are no more than several times $\omega/c$, but at the same time no shorter than $\omega/c$. In this way a finite dimensional linear system is obtained that can be solved for $r(\pvec{k}+\pvec{G}(h)|\pvec{k})$ by standard methods.

\section{Diffraction Efficiencies}
The total time-averaged flux scattered from our doubly-periodic surface is obtained by substituting Eq.~\eqref{eq:24} into Eq.~\eqref{eq:15}:
\begin{align}
  \label{eq:33}
  P_{\mathrm{sc}}
  &=
    A
    \int\limits_{q_\parallel<\omega/c}
    \dfint[2]{q_\parallel}{(2\pi)^2}
    \alpha_0(q_\parallel,\omega)
    \sum\limits_{\pvec{G}}
    (2\pi)^2 \delta\left( \pvec{q} - \pvec{k} - \pvec{G} \right)
    r^*( \pvec{k} + \pvec{G}  | \pvec{k} )
    \notag \\
  &
    \qquad \qquad \qquad \qquad \times 
    \sum\limits_{\pvec{G}'}
    (2\pi)^2 \delta\left( \pvec{q} - \pvec{k} - \pvec{G}' \right)
    r( \pvec{k} + \pvec{G}' | \pvec{k} ).
\end{align}
The only nonzero terms on the right-hand side of this equation are those for which $\pvec{G}'=\pvec{G}$. Then, with the result that in two-dimensions
\begin{align}
  \label{eq:34}
  \left( 2\pi \right)^2 \delta(\vec{0}) &= L_1 L_2,
\end{align}
Eq.~\eqref{eq:33} becomes
\begin{align}
  \label{eq:35}
    P_{\mathrm{sc}}
  &=
    A L_1 L_2
    \int\limits_{q_\parallel<\omega/c}
    \dint[2]{q_\parallel}
    \alpha_0(q_\parallel,\omega)
    \sum\limits_{\pvec{G}}
    \delta\left( \pvec{q} - \pvec{k} - \pvec{G} \right)
    \left| r( \pvec{k} + \pvec{G} | \pvec{k} ) \right|^2
    \notag \\
  &=
    A L_1 L_2
    \left. \sum\limits_{\pvec{G}} \right.^\prime 
    \alpha_0(|\pvec{k}+\pvec{G}|, \omega )
    \left| r(\pvec{k} + \pvec{G} | \pvec{k} ) \right|^2.
\end{align}
The prime on the sum indicates that it extends over only those values of $\pvec{G}$ for which $|\pvec{k}+\pvec{G}|<\omega/c$. Equation~\eqref{eq:35} demonstrates that each diffracted beam contributes independently to the scattered flux.

When the scattered flux is normalized by the total time-averaged flux of the incident field, Eq.~\eqref{eq:13}, the result can be written
\begin{align}
  \label{eq:36}
  \frac{ P_{\mathrm{sc}} }{ P_{\mathrm{inc}}  }
  &=
    \left. \sum\limits_{\pvec{G}} \right.^\prime 
    e\left( \pvec{k}+\pvec{G} | \pvec{k} \right),
\end{align}
where
\begin{align}
  \label{eq:37}
  e \left( \pvec{k}+\pvec{G} | \pvec{k} \right)
  &=
    \frac{
    \alpha_0(|\pvec{k}+\pvec{G}|, \omega)
    }{
    \alpha_0(k_\parallel, \omega )
    }
    \,
    \left| r( \pvec{k}+\pvec{G}|\pvec{k}) \right|^2.
\end{align}
The quantity $e \left( \pvec{k}+\pvec{G} | \pvec{k} \right)$, called the \textit{diffraction efficiency}, is the fraction of the total time-averaged incident flux that is diffracted into a Bragg beam defined by the wave vector $\pvec{k}+\pvec{G}$ (when the incident beam is defined by $\pvec{k}$). It has a physical meaning for only those values of $\pvec{G}$ for which $\alpha_0(|\pvec{k}+\pvec{G}|, \omega)$  is real. The propagating diffracted beams defined by this condition are called the \textit{open channels}.

Since there is no absorption in the scattering from an impenetrable surface, all the power incident on it must be scattered back into the medium of incidence. Hence, the conservation of energy in the scattering process requires that
\begin{align}
  \label{eq:38}
    \left. \sum\limits_{\pvec{G}} \right.^\prime 
  e\left( \pvec{k}+\pvec{G} | \pvec{k} \right)
  &=
    1.
\end{align}
The closeness to unity of the sum on the left-hand side of Eq.~\eqref{eq:38} is a good test of the quality of the numerical simulation calculations of the diffraction efficiencies.

The \textit{reflectivity} of the bigrating is obtained from the diffraction efficiency for the beam defined by $\pvec{G}=\vec{0}$:
\begin{align}
  \label{eq:39}
  \mathcal{R}(\pvec{k})
  &=
    e(\pvec{k}|\pvec{k}).
\end{align}





\section{Results}

We will illustrate the proceeding results by presenting simulation results for the dependence of the reflectivity and several other diffraction efficiencies on the polar and azimuthal angles of incidence $\theta_0$ and $\phi_0$, respectively, when the bigrating defined by the surface profile function 
\begin{align}
  \label{eq:cosine_surface_profile}
  \zeta(\pvec{x})
  &=
    \frac{\zeta_0}{2}
    \left[
    \cos\left( \frac{2\pi x_1}{a} \right)
    +
    \cos\left( \frac{2\pi x_2}{a} \right)
    \right]
\end{align}
is illuminated by a scalar plane wave of frequency $\omega$. The primitive translation vectors of the square Bravais lattice underlying this surface profile function are
\begin{align}
  \vec{a}_1 &= a (1,0,0)
  &
    \vec{a}_2 &= a (0,1,0).
\end{align}
Those of the corresponding reciprocal lattice are
\begin{align}
  \vec{b}_1 &= \frac{2\pi}{a} (1,0,0)
  &
    \vec{b}_2 &= \frac{2\pi}{a} (0,1,0).
\end{align}

An attractive feature of the form of the surface profile function in Eq.~\eqref{eq:cosine_surface_profile} is that the $\widehat{I}$-integral defined in Eq.~\eqref{eq:29} can be obtained analytically, and takes the form
\begin{align}
  \widehat{I}\left(\gamma| \pvec{G}(h) \right)
  &=
    (-1)^{h_1+h_2}
    \operatorname{J}_{h_1}\!\!\left( \frac{\gamma \zeta_0}{2} \right)
    \operatorname{J}_{h_2}\!\!\left( \frac{\gamma \zeta_0}{2} \right),
\end{align}
where $\operatorname{J}_n (\cdot)$ represents the Bessel function of the first kind and order $n$. 

\begin{figure}[tb]
   \centering \includegraphics[width=0.65\linewidth]{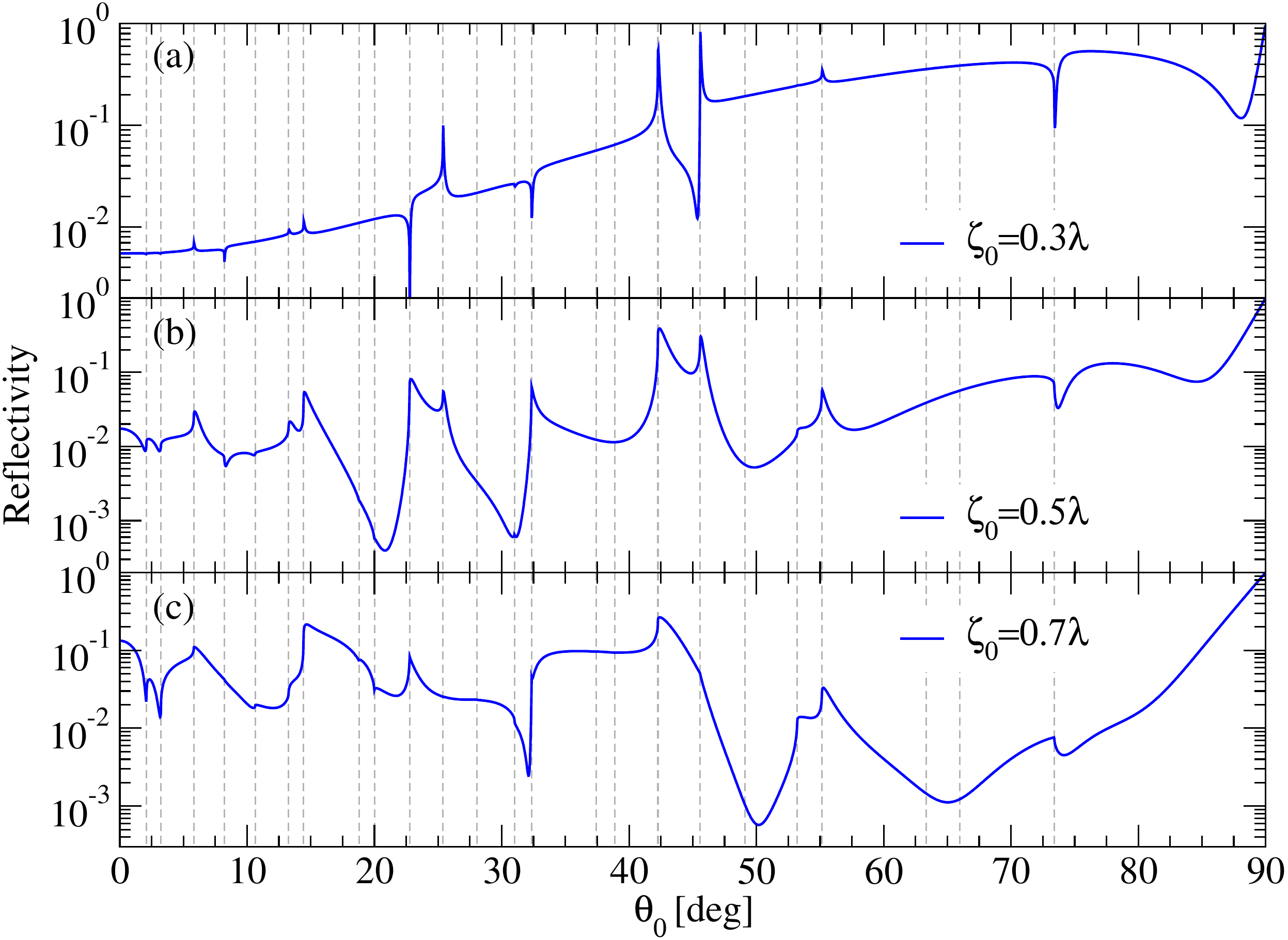}
   \caption{Reflectivity of a doubly-periodic cosine Neumann surface [see Eq.~\eqref{eq:cosine_surface_profile}] as a function of the polar angle of incidence $\theta_0$ for the  azimuthal angle of incidence $\phi_0=\ang{0}$. The doubly-periodic cosine grating had a period $a/\lambda=\num{3.5}$ and amplitudes (a)~$\zeta_0/\lambda=0.3$; (b)~$\zeta_0/\lambda=0.5$ and (c)~$\zeta_0/\lambda=0.7$. The vertical dashed lines display the positions of the Rayleigh anomalies predicted on the basis of Eq.~\eqref{eq:46}. The scan over polar angle of incidence, $\theta_0$,  was done in steps of $\Delta \theta_0=\ang{0.025}$. In performing the numerical calculations, it was assumed that $G_\parallel(h)\leq 4\omega/c$.}
   \label{Fig:01}
 \end{figure}

In the first set of calculations of the dependence of the reflectivity of the bigrating defined by Eq.~\eqref{eq:cosine_surface_profile} on the polar angle of incidence $\theta_0$ for a value of the azimuthal angle of incidence $\phi_0=\ang{0}$ [$\pvecUnit{k}=(1,0,0)$],  we assumed that the lattice constant $a$ had the value $a=3.5\lambda$, where $\lambda$ is the wavelength of the incident wave, while the amplitude $\zeta_0$ took several values. The  calculated reflectivities for Neumann surfaces characterized by the amplitudes $\zeta_0=0.3\lambda$, $0.5\lambda$, $0.7\lambda$ are presented in Fig.~\ref{Fig:01}. These results show a complex dependence of the reflectivity on the polar angle of incidence in the form of the presence of many sharp peaks and dips. These features are  \emph{Rayleigh anomalies}, which occur at values of $\theta_0$ for a given value of $\phi_0$ at which diffractive orders start or cease to propagate.

%

To determine the angles of incidence at which the Rayleigh anomalies occur, we note that the lateral wave vector of the diffractive beam characterized by the index pair ${h_1, h_2}$ is given by
\begin{align}
  \label{eq:44}
  \pvec{q}(h_1,h_2) &= \pvec{k} + \pvec{G}(h_1,h_2).
\end{align}
This diffractive beam goes from being a propagating one to an evanescent one when $|\pvec{q}(h_1,h_2)|=\omega/c$, which is the condition for a potential Rayleigh anomaly to be associated with this wave.  On squaring both sides of Eq.~\eqref{eq:44} and using Eqs.~\eqref{eq:16a} and \eqref{eq:22}, we obtain a quadratic equation for $\sin\theta_0$
\begin{align}
  \label{eq:45}
  \sin^2\theta_0
  +
  2 \sin\theta_0 \,
  \pvecUnit{k} \cdot
  \left(
  h_1\frac{c}{\omega} \vec{b}_1
  +
  h_2\frac{c}{\omega} \vec{b}_2
  \right)
  +
  \left(
  h_1\frac{c}{\omega} \vec{b}_1
  +
  h_2\frac{c}{\omega} \vec{b}_2
  \right)^2 -1
  &=
    0,
\end{align}
with $\pvecUnit{k}=(\cos\phi_0,\sin\phi_0,0)$.

Equation~\eqref{eq:45} determines for a general grating where Rayleigh anomalies can exist. From its solutions 
\begin{align}
  \label{eq:46}
  \sin\theta_0
  &=
    - \pvecUnit{k} \cdot
    \left(
    h_1\frac{c}{\omega} \vec{b}_1
    +
    h_2\frac{c}{\omega} \vec{b}_2
    \right)
    \pm
    \left\{
    \left[
    \pvecUnit{k} \cdot
    \left(
    h_1\frac{c}{\omega} \vec{b}_1
    +
    h_2\frac{c}{\omega} \vec{b}_2
    \right)
    \right]^2
    -
    \left[
    h_1\frac{c}{\omega} \vec{b}_1
    +
    h_2\frac{c}{\omega} \vec{b}_2
    \right]^2
    +
    1
    \right\}^\frac{1}{2},
\end{align}
under the condition $|\sin\theta_0| \leq 1$, as $h_1$ and $h_2$ each run over all positive and negative integers and zero, the polar angles of incidence $\theta_0$ at which Rayleigh anomalies can exist for a specified azimuthal angle of incidence $\phi_0$ are obtained. The values of $\theta_0$ obtained in this way are indicated by gray vertical dashed lines in Fig.~\ref{Fig:01}.  From the results of this figure we see that the majority of the peaks and dips present in the refelctivity are Rayleigh anomalies. It should be noted that even if a Rayleigh anomaly is predicted to exist at a particular polar angle of incidence, it may not be observed in the reflectivity, because it is too weak to be seen.

We see from Fig.~\ref{Fig:01} that as the amplitude of the surface profile function $\zeta_0$ is increased, the polar angles of incidence at which the Rayleigh anomalies occur do not change, as must be the case, but the forms of the anomalies can change. Peaks and dips can change their magnitudes, and broaden, and dips can change into peaks, and peaks can change into dips.

The numerical calculations that produced the results presented in Fig.~\ref{Fig:01} were performed under the assumption that $G_\parallel(h)\leq 4\omega/c$, and the linear system of equations satisfied by $r(\pvec{K}|\pvec{k})$, Eq.~\eqref{eq:32}, was solved by the routine \textrm{la}\verb!_!\textrm{gesv} from LAPACK95~\cite{Book:Barker2001}. For this  value of $\max
  G_\parallel(h)$ the simulation time required per angle of incidence to produce the results in Fig.~\ref{Fig:01} was \SI{1.5}{s}, or less on average,  when the simulations were performed on a machine equipped with an Intel i7-5930K CPU running at \SI{3.50}{GHz}. The energy conservation condition~\eqref{eq:38} was found to be satisfied with an error no greater than \num{E-10} for all the values of $\theta_0$ and $\zeta_0$ that we considered.

%
 \begin{figure}[tb]
   \centering \includegraphics[width=0.65\linewidth]{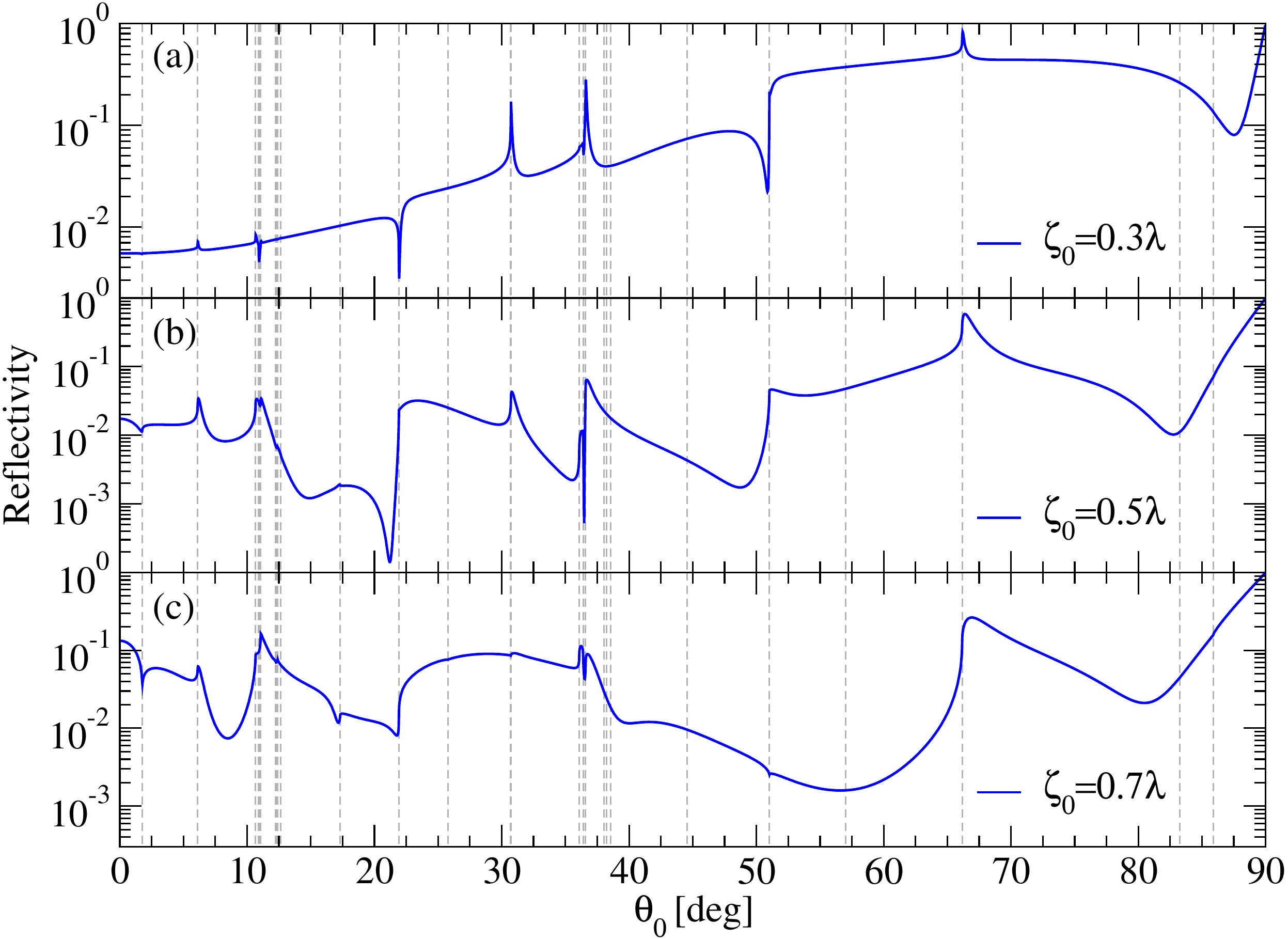}
   \caption{The same as Fig.~\protect\ref{Fig:01} but for $\phi_0=\ang{45}$.}   \label{Fig:01-phi0=45}
 \end{figure}

 In Fig.~\ref{Fig:01-phi0=45} we present the dependence of the reflectivity on the polar angle of incidence when the azimuthal angle of incidence is $\phi_0=\ang{45}$. In this case the unit vector $\pvecUnit{k}$ becomes $\pvecUnit{k}=(1/\sqrt{2},1/\sqrt{2},0)$, and the values of $\theta_0$ at which Rayleigh anomalies are predicted to occur are different from the values at which they occur in Fig.~\ref{Fig:01}. With an increase of the amplitude of the surface profile function, these anomalies undergo the same kinds of changes in their forms as they do in the case where $\phi_0=\ang{0}$.

\begin{figure}[b]
   \centering \includegraphics[width=0.65\linewidth]{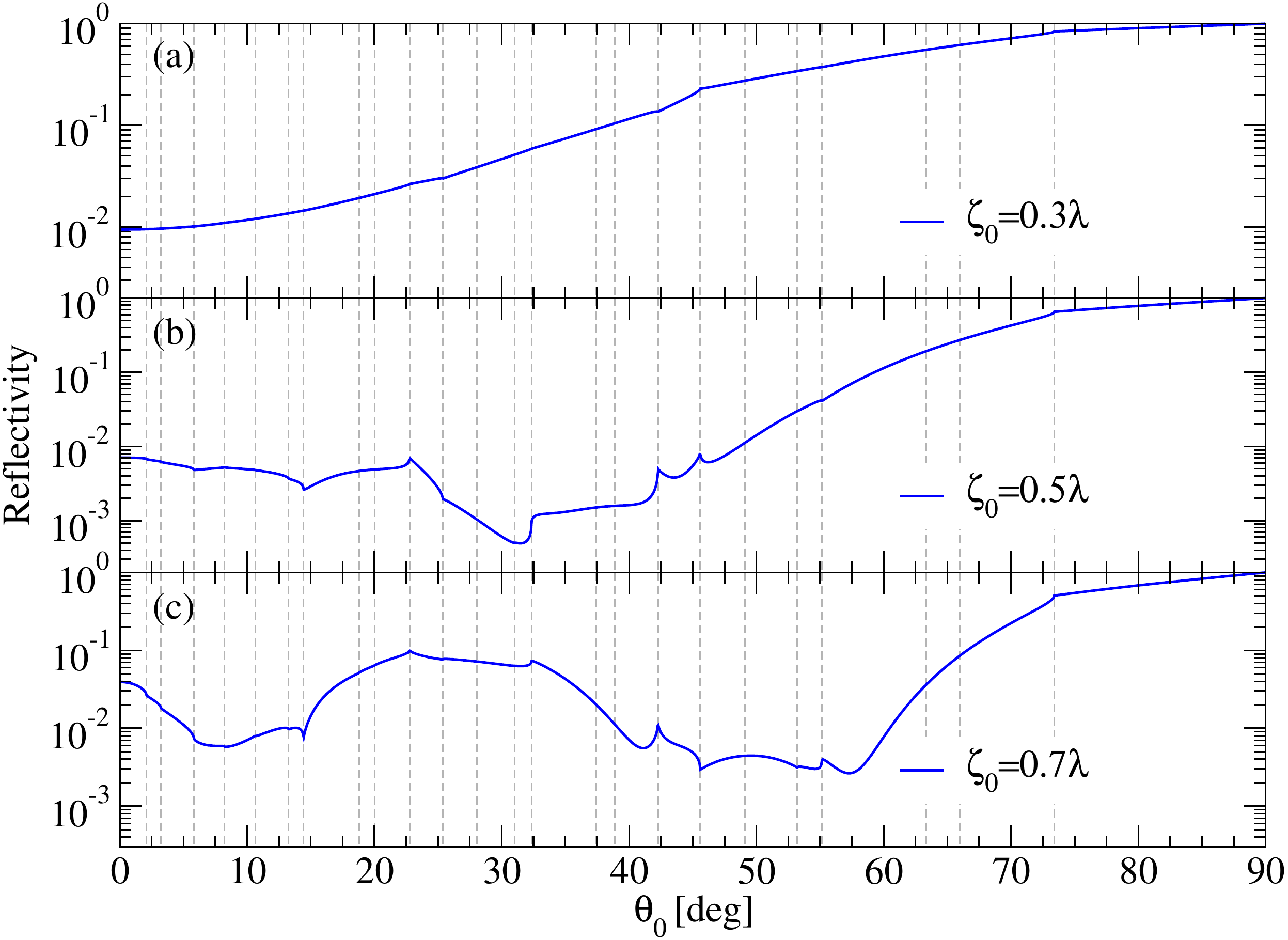} 
   \caption{Same as Fig.~\protect\ref{Fig:01} but for doubly-periodic Dirichlet surfaces.}
   \label{Fig:02}
 \end{figure}

\medskip
We now turn to the diffraction of a scalar plane wave from a doubly-periodic Dirichlet surface defined by Eq.~\eqref{eq:cosine_surface_profile}. In Fig.~\ref{Fig:02} we present the reflectivity as a function of $\theta_0$ for the case where $\phi_0=\ang{0}$. The parameters defining the surface profile function are $a=3.5\lambda$ and $\zeta_0=0.3\lambda$, $0.5\lambda$, $0.7\lambda$, namely the values assumed in obtaining the results presented in Fig.~\ref{Fig:01} and \ref{Fig:01-phi0=45}. The values of $\theta_0$ at which Rayleigh anomalies are predicted to exist are the same as those at which they are predicted to exist in Fig.~\ref{Fig:01}. However, these anomalies are significantly weaker than those occurring at the same values of $\theta_0$ in Fig.~\ref{Fig:01}. This difference demonstrates the important role  played by the boundary condition on the surface of the bigrating satisfied by the field in the region $x_3>\zeta(\pvec{x})$ in forming these anomalies.

%

%
\medskip
A further comparison of Figs.~\ref{Fig:01} and \ref{Fig:02} prompts the following observation. In Fig.~\ref{Fig:01}(a) we see a dip in the reflectivity at a value of $\theta_0\approx \ang{88.0}$, an angle at which no Rayleigh anomaly is predicted to exist. In Fig.~\ref{Fig:01}(b), for a larger amplitude of the bigrating profile function, this dip has broadened and shifted to a smaller value of $\theta_0$, namely $\theta_0\approx \ang{84.5}$. Again, no Rayleigh anomaly is predicted to occur at this angle. With a further increase of the amplitude of the bigrating profile function, we see in Fig.~\ref{Fig:01}(c) a break in the slope of the reflectivity curve at a value of $\theta_0\approx \ang{80}$. Such a dip is more clearly visible at these three values of $\theta_0$ in Figs.~\ref{Fig:01-phi0=45}(a)--(c). No such feature is present at these (or other) angles in Figs.~\ref{Fig:02}(a)--(c). It is known~\cite{14} that a doubly-periodic Neumann surface supports a surface wave, while a doubly-periodic Dirichlet surface does not. It is also known that changing the amplitude of the surface profile function shifts the nonradiative and radiative  branches of the dispersion relation (in the reduced zone scheme) of the surface wave on a Neumann bigrating~\cite{14}. Since a Wood anomaly arises due to the excitation of a surface wave on a periodically modulated surface by the incident field~\cite{22,23} the angles of incidence at which these anomalies occur will shift with changes in the surface profile function. These properties of the large angle dip suggest that it represents a Wood anomaly. However, confirmation of this conjecture has to await the determination of the branches of the dispersion curve of the surface wave supported by the doubly-periodic Neumann surface, in the radiative region of the $(\pvec{k},\omega)$ plane as well as in the nonradiative region.

%

%
 \begin{figure}[b]
   \centering \includegraphics[width=0.65\linewidth]{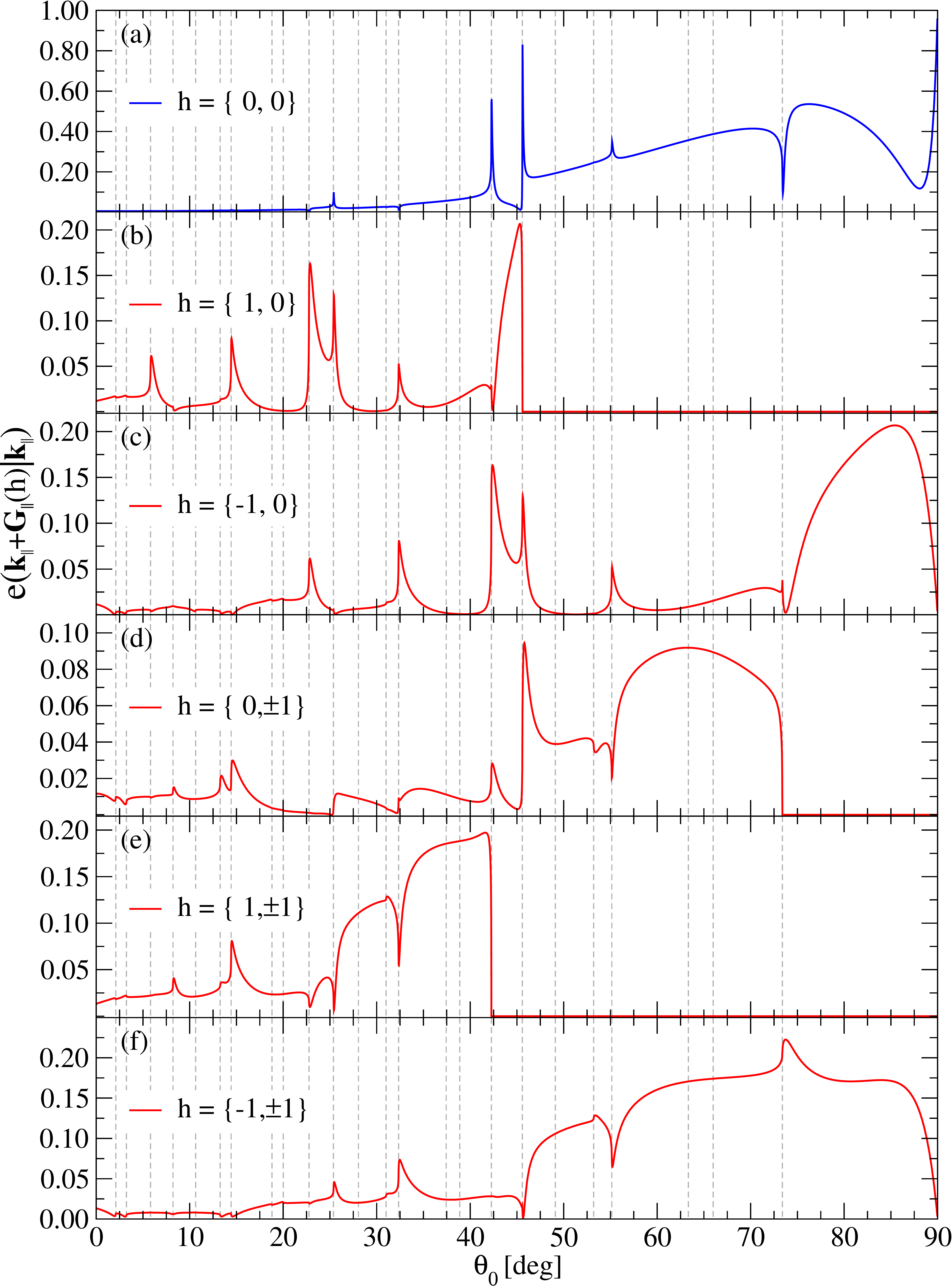}
   \caption{Several diffraction efficiencies $e(\pvec{k}+\pvec{G}(h)|\pvec{k})$ for values of $h$  given in each panel as functions of the polar angle of incidence $\theta_0$ for the  azimuthal angle of incidence $\phi_0=\ang{0}$. The doubly-periodic Neumann surface, defined by Eq.~\eqref{eq:cosine_surface_profile}, is characterized by the parameters $\zeta_0=\num{0.5}\lambda$  and $a=\num{3.5}\lambda$. These are the same parameter values assumed in obtaining the results presented in  Figs.~\ref{Fig:01}(b), \ref{Fig:01-phi0=45}(b), and \ref{Fig:02}(b). Here also the scan over the polar angle of incidence was done in steps of $\Delta \theta_0=\ang{0.025}$, and $G_\parallel(h)\leq 4\omega/c$ was assumed in performing the numerical calculations.}
\label{Fig:01-efficiencies}
 \end{figure}

Features similar to those observed in Figs.~\ref{Fig:01}--\ref{Fig:02} are present in the dependence of other diffraction efficiencies on $\theta_0$ for a given value of $\phi_0$. In Fig.~\ref{Fig:01-efficiencies} we plot this dependence for the efficiencies of the $\{1,0\}$, $\{-1,0\}$, $\{0,\pm 1\}$, $\{1,\pm 1\}$, and $\{-1,\pm 1\}$ beams diffracted from the Neumann surface defined by Eq.~\eqref{eq:cosine_surface_profile} with $\zeta_0=\num{0.5}\lambda$ and $a=\num{3.5}\lambda$. The azimuthal angle of incidence is $\phi_0=\ang{0}$. The notation $\{h_1,\pm h_2\}$ indicates that the  $\{h_1,h_2\}$ and $\{h_1,-h_2\}$ efficiencies are identical. This identity is a consequence of the symmetry of the scattering system under reflection in the $x_1$ axis when $\phi_0=\ang{0}$. The predicted angular positions of the Rayleigh anomalies are indicated by the gray vertical dashed lines. It is seen that all of the peaks and dips in these dependencies occur at these angles, but not every one of these angles has an anomaly associated with it.

\begin{figure}[b]
    \centering
    \includegraphics[height=0.40\linewidth]{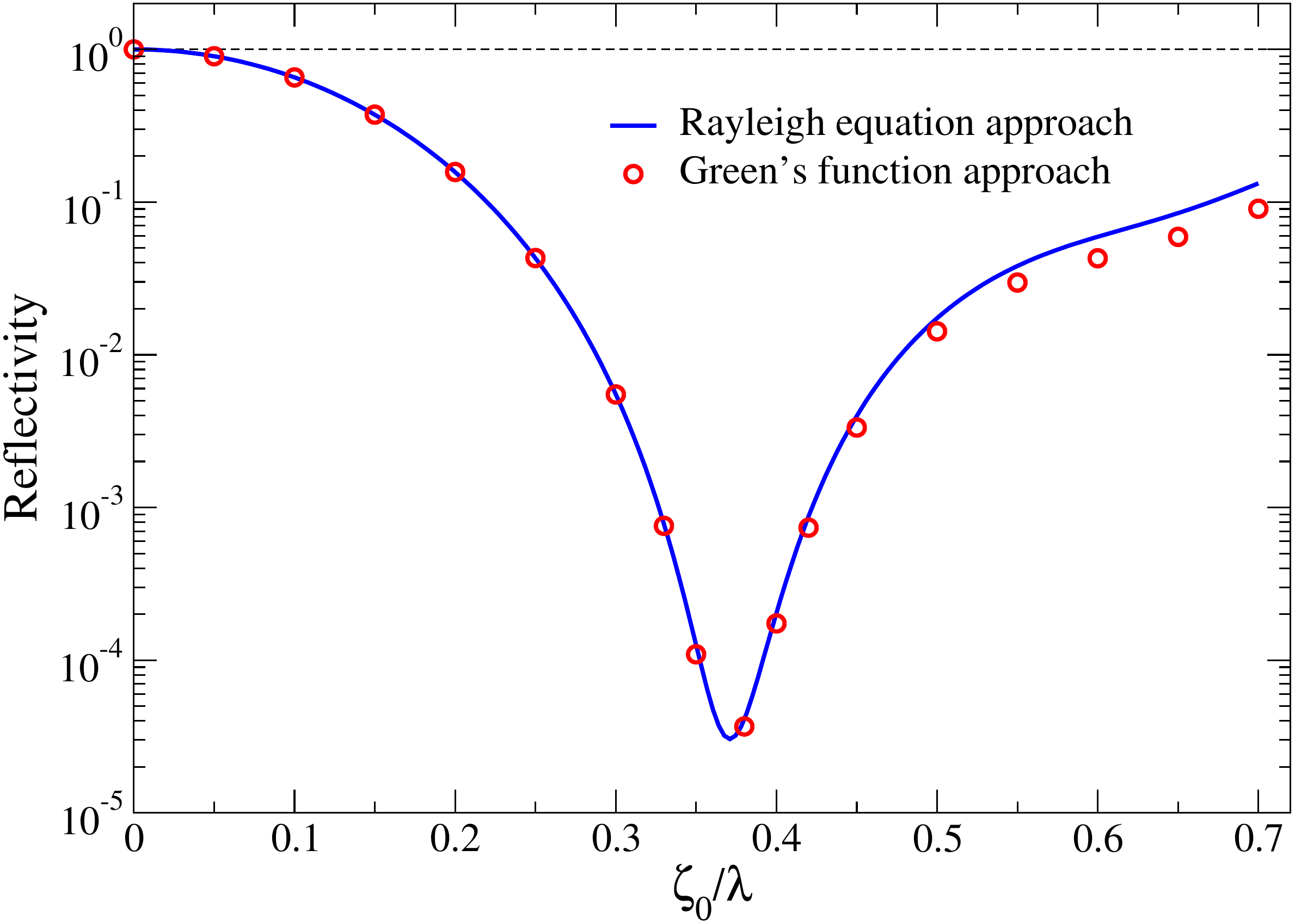} 
    \caption{Reflectivity of a doubly-periodic cosine Neumann surface [see Eq.~\eqref{eq:cosine_surface_profile}] of period $a/\lambda = 3.5$ as a function of the amplitude $\zeta_0$ for the polar and azimuthal angles of incidence $\theta_0=\ang{0}$ and   $\phi_0=\ang{0}$, respectively.
      The solid line represents the results obtained on the basis of the Rayleigh equation~\eqref{eq:32} while the open symbols were obtained by a rigorous Green's function-based numerical approach~\cite{17}. In performing the latter calculations it was assumed the edge of the square region of the $x_1x_2$ plane covered by the  doubly-periodic surface was $L=38\lambda$, the width of the Gaussian incident beam was $W=15\lambda$, and $\Delta x=0.15\lambda$ was the discretization interval used.}
    \label{Fig:05}
  \end{figure}

%
%
 \medskip
 It is apparent from the results presented  in Fig.~\ref{Fig:01}, for instance,  that the reflectivity of the doubly-periodic cosine Neumann surface depends strongly on its amplitude $\zeta_0$. To further investigate this dependence, we present in Fig.~\ref{Fig:05} as a solid line the reflectivity of such a surface of period $a=3.5\lambda$ as a function of the amplitude $\zeta_0$ for polar and azimuthal angles of incidence $\theta_0=\ang{0}$ and   $\phi_0=\ang{0}$, respectively. These results were obtained on the basis of a numerical solution of the Rayleigh equation~\eqref{eq:32} for the same values of the numerical parameters assumed in obtaining the results in Fig.~\ref{Fig:01}. Figure~\ref{Fig:05} shows that the reflectivity of the doubly-periodic  Neumann surface decreases monotonically from unity to approximately \num{3E-5} when its amplitude increases from zero (planar surface) to $\zeta_0=\num{0.371}\lambda$. Increasing the amplitude beyond this value causes the reflectivity of the surface to increase monotonically, and it reaches the value \num{0.1357} when $\zeta_0=0.7\lambda$. What happens to the reflectivity when $\zeta_0/\lambda>0.7$, we have not investigated here.
  
%
%
 To validate our use of the Rayleigh equation in obtaining the results presented in this work, we performed additional calculations obtained on the basis of a rigorous Green's function-based numerical approach~\cite{17,17a}. To this end, the latter approach was used to calculate the reflectivity for normal incidence
 as a function of the corrugation strength $\zeta_0$. The results of such calculations are presented as open symbols in Fig.~\ref{Fig:05}, and they show satisfactory agreement with the corresponding results obtained on the basis of the Rayleigh equation approach. In particular, the five orders of magnitude variation of the reflectivity is consistently predicted by both approaches. It is only for $\zeta_0/\lambda>0.5$ that  some minor discrepancy starts to develop. As we will comment below, it it not entirely clear if this should be interpreted an indication that the Rayleigh equation approach starts to become less accurate.
 
  We now briefly detail how the rigorous Green's function-based numerical calculations were performed; reference~\onlinecite{17a} gives additional details. Since the Green's function-based approach as formulated in Ref.~\onlinecite{17a} does not explicitly use the fact that the surface is periodic, the first step of the calculation is to restrict the doubly-periodic  surface~\eqref{eq:cosine_surface_profile} to a square region of the $x_1x_2$ plane of edges $L$. Next, this surface profile, as well as its derivatives up to order two, are discretized on a square lattice of points of lattice parameters $\Delta x$. To avoid diffraction artifacts from the edges of the surface, the incident beam is assumed to be a Gaussian beam of $1/\mathrm{e}$ half-width $W$. In the numerical calculations using the Green's function approach that we report in Fig.~\ref{Fig:05} the values of the numerical parameters were $L=38\lambda$,  $W=15\lambda$, and $\Delta x =0.15 \lambda$. The reflectivity of the surface was calculated by integrating the differential reflection coefficient $\partial R /\partial\Omega_s$ over an angular region around the specular direction in such a way that only the contribution from the fundamental diffractive order was included. Since the period of the surface that we consider is sufficiently long [$a=3.5\lambda$], the diffractive orders were well separated. For the calculation of the reflectivity we used a region defined by $|\pvec{q}-\pvec{k}|<0.1\,\omega/c$. We also checked and found that minor adjustments of the size of this angular region did not affect the reflectivity values obtained in any significant way. It should be remarked that
for the largest values of $\zeta_0$ that we considered we found a weak but  detectable dependence of the reflectivity on the parameters $L$ and $W$ when it was calculated by the Green's function approach as described above. Hence the discrepancy seen in Fig.~\ref{Fig:05}  in the reflectivity obtained by the two approaches for the largest values of the corrugation strength is not necessarily due to the Rayleigh approach becoming inaccurate.

The calculations based on the  Green's function approach, whose results are reported in Fig.~\ref{Fig:05}, took about \SI{34}{min} (or about \SI{2000}{s}) of cpu-time to complete per value of $\zeta_0$ when the calculation was performed on an Intel i7 960 processor running at \SI{3.20}{GHz}, and the memory footprint of the calculation was almost \SI{19}{Gb}. A similar calculation using the Rayleigh equation approach took about \SI{1}{s} of cpu-time when performed on the same computer for the numerical parameters that we assumed and it required only a faction of the computer memory needed by the Green's function calculation.

Finally we should remark that the rigorous Green's function approach~\cite{17a} described and used above is not ideal for a doubly-periodic system. An approach of this kind that is adapted to doubly-periodic systems uses periodic Green's functions~\cite{26}. However, the usual expressions for doubly-periodic Green's functions contain slowly convergent series~\cite{26}, and have to be subjected to accelerated transformations~\cite{26,27,28} to make them useful in calculations. We have therefore decided not to pursue it in this work.

\section{Conclusions}
We have derived the Rayleigh equation for the amplitude of the scattered field when a scalar plane wave is incident on a two-dimensional rough surface on which the Neumann or Dirichlet boundary condition is imposed. From this equation we have obtained the equation for the amplitudes of the diffracted Bragg beams, when the rough surface is a doubly-periodic one. This equation has been solved by a rigorous numerical approach, and from the solution the dependence of the  diffraction efficiencies of several of the of the lowest-order diffracted beams on the polar and azimuthal angles of incidence has been determined. These dependencies display a rich structure of peaks and dips as functions of the polar angle of incidence for a specified azimuthal angle of incidence. These features occur at the angles at which a diffracted beam starts or ceases to propagate. Hence they are the analogues for a doubly-periodic grating of the anomalies that were first observed by Wood~\cite{15} in the diffraction of light by a classical metal grating, and were subsequently explained by Lord~Rayleigh~\cite{16}. They are now called Rayleigh anomalies. These anomalies are observed in the diffraction of a scalar wave from both a Neumann and a Dirichlet surface. In the case of diffraction from a Neumann surface an additional anomaly, a dip, is observed in the reflectivity at angles of incidence for which no Rayleigh anomaly is predicted to occur.  No such anomaly is presented in diffraction from doubly-periodic Dirichlet surfaces. A doubly-periodic Neumann surface supports a surface wave, while a doubly-periodic Dirichlet surface does not. From this and responses of the dip to changes of the surface profile function of the Neumann surface, it is conjectured that it is a Wood anomaly that was first reported by Wood in Ref.~\onlinecite{15}, and was subsequently explained by Fano~\cite{22} as due to the excitation of the surface electromagnetic wave supported by the grating by the incident light through the periodic modulation of the surface. This conjecture can only be verified when the branches of the dispersion curve of the surface wave on the Neumann bigrating in the radiative region of the $(\pvec{k},\omega)$ plane have been determined. That will be the subject of a separate work. It should be noted that neither the Neumann nor the Dirichlet surface supports a surface wave when it is planar.
Finally, by comparing results obtained from solutions of the Rayleigh equation with results obtained by a rigorous Green's function-based numerical approach~\cite{17}, we have validated the use of the Rayleigh equation in the calculations reported here.

\begin{acknowledgments}
We dedicate this paper to the memory of Arnold Markovich Kosevich. He contributed outstanding work to many areas of condensed matter theory, and it was a pleasure to know him. The research of I.S. was supported in part by the Research Council of Norway~(Contract 216699) and the French National Research Agency (ANR-15-CHIN-0003).
\end{acknowledgments}

\appendix
\section{The mean differential reflection coefficient}

%
For completeness we note that if the surface profile function $\zeta(\pvec{x})$ is a stationary, zero-mean, isotropic random process, it is the average of the differential reflection coefficient over the ensemble of realizations of $\zeta(\pvec{x})$ that we must calculate:
\begin{align}
  \label{eq:a}
  \left< \frac{\partial R }{ \partial \Omega_s} \right>
  &=
    \frac{1}{L_1 L_2}
    \left( \frac{\omega}{2\pi c}\right)^2
    \frac{ \cos^2\theta_s }{ \cos\theta_0 }
    \left< \left| R(\pvec{q}|\pvec{k}) \right|^2 \right>,        
\end{align}
Here, and in all that follows, the angle brackets denote an average over the ensemble of realizations of the surface profile function. If we write the scattering amplitude $R(\pvec{q}|\pvec{k})$ as the sum of its average value and the fluctuation from the mean value
\begin{align}
  \label{eq:b}
  R(\pvec{q}|\pvec{k})
  &=
    \left< R(\pvec{q}|\pvec{k}) \right>
    +
    \left[
    R(\pvec{q}|\pvec{k}) - \left< R(\pvec{q}|\pvec{k}) \right>
    \right],
\end{align}
we find that each term contributes separately to the mean differential reflection coefficient,
\begin{align}
  \label{eq:c}
  \left< \frac{\partial R }{ \partial \Omega_s} \right>
  &=
    \left< \frac{\partial R }{ \partial \Omega_s} \right>_{\mathrm{coh}}
    +
    \left< \frac{\partial R }{ \partial \Omega_s} \right>_{\mathrm{incoh}}
\end{align}
where
\begin{align}
  \label{eq:d}
  \left< \frac{\partial R }{ \partial \Omega_s} \right>_{\mathrm{coh}}
  &=
    \frac{1}{L_1 L_2}
    \left( \frac{\omega}{2\pi c}\right)^2
    \frac{ \cos^2\theta_s }{ \cos\theta_0 }
    \left| \left< R(\pvec{q}|\pvec{k}) \right> \right|^2
\end{align}
and
\begin{align}
  \label{eq:e}
  \left< \frac{\partial R }{ \partial \Omega_s} \right>_{\mathrm{incoh}}
  &=
    \frac{1}{L_1 L_2}
    \left( \frac{\omega}{2\pi c}\right)^2
    \frac{ \cos^2\theta_s }{ \cos\theta_0 }
    \left< \left|
    R(\pvec{q}|\pvec{k}) - \left< R(\pvec{q}|\pvec{k}) \right>
    \right|^2 \right>
    \notag \\
  &=
    \frac{1}{L_1 L_2}
    \left( \frac{\omega}{2\pi c}\right)^2
    \frac{ \cos^2\theta_s }{ \cos\theta_0 }
    \left[
    \left<
    \left| R(\pvec{q}|\pvec{k}) \right|^2\right>
    -
    \left| \left< R(\pvec{q}|\pvec{k})  \right> \right|^2
    \right].
\end{align}
The first term on the right-had side of Eq.~\eqref{eq:c} is the contribution to the mean differential reflection coefficient from the field scattered coherently (specularly), while the second term is the contribution from the field scattered incoherently (diffusely). Recently, expressions similar to those that appear in Eqs.~\eqref{eq:d} and \eqref{eq:e} were used to calculate the mean differential reflection coefficient on the basis of the numerical solutions of the Rayleigh equations for the scattering of light from a two-dimensional randomly rough perfectly conducting surface~\cite{PEC}.

For the type of randomly rough surfaces considered here  it is the case that
\begin{align}
  \label{eq:f}
  \left< R(\pvec{q}|\pvec{k}) \right>
  &=
    (2\pi)^2 \delta(\pvec{q}-\pvec{k}) r(k_\parallel).
\end{align}
The delta function is a consequence of the assumed stationary of the surface profile function, while the fact that $r(k_\parallel)$ is a function of $k_\parallel$ only through its magnitude is due to the isotropy of the surface profile function.

The reflectivity of the randomly rough surface is given by
\begin{align}
  \label{eq:g}
  \mathcal{R}(\theta_0)
  &=
    \int\limits_0^{\frac{\pi}{2}} \dint{\theta_s} \sin\theta_s
    \int\limits_{-\pi}^{\pi} \dint{\phi_s}
    \left< \frac{\partial R }{ \partial \Omega_s} \right>_{\mathrm{coh}}.    
\end{align}
With the use of Eqs.~\eqref{eq:d}, \eqref{eq:f}, \eqref{eq:34} and the result that
\begin{align}
  \label{eq:h}
  \delta\left( \pvec{q} - \pvec{k} \right)
  &=
    \left( \frac{c}{\omega}\right)^2
    \frac{
    \delta(\theta_s-\theta_0)  \delta(\phi_s-\phi_0) 
    }{
    \sin\theta_0  \cos\theta_0
    },
\end{align}
Eq.~\eqref{eq:g} simplifies to
\begin{align}
  \label{eq:i}
  \mathcal{R}(\theta_0)
  &=
    \left| r(k_\parallel) \right|^2
    =
    \left| r\left( \frac{\omega}{c} \sin\theta_0 \right) \right|^2.
\end{align}
From Eqs.~\eqref{eq:f} and \eqref{eq:34} we find that
\begin{align}
  \label{eq:j}
  r(k_\parallel)
  &=
    \frac{1}{L_1 L_2}
    \left< R(\pvec{k}|\pvec{k}) \right>.
\end{align}

%
\bibliography{paper2017-06} 

\begin{thebibliography}{39}
\bibitem [{21()}]{21}%
  \BibitemOpen
  \href@noop {} {}\bibinfo {note} {Equation~(27) is obtained by applaying the
  two-dimensional version of the Poisson summation formula to the sum on the
  left-hand side of it.}\BibitemShut {Stop}%
\end{thebibliography}

\end{document}